\documentclass{llncs}
\usepackage{todonotes}
\usepackage{amsmath}
\usepackage{graphicx}
\usepackage{hyperref}
\usepackage{algorithm} 
\usepackage{algpseudocode}
\begin{document}

\title{Semantic Place Descriptors for Classification and Map Discovery}

\author{Siddharth Sarda\inst{1} \and Carsten Eickhoff\inst{2} \and Thomas Hofmann\inst{2}}
\institute{Booking.com, Amsterdam, The Netherlands\\
\email{x@booking.com}
\and
ETH Zurich, Dept.\ of Computer Science, Zurich, Switzerland\\
\email{\{carsten.eickhoff, thomas.hofmann\}@inf.ethz.ch}
}

\maketitle

\begin{abstract}
Urban environments develop complex, non-obvious structures that are often hard to represent in the form of maps or guides. Finding the right place to go often requires intimate familiarity with the location in question and cannot easily be deduced by visitors. In this work, we exploit large-scale samples of usage information, in the form of mobile phone traces and geo-tagged Twitter messages in order to automatically explore and annotate city maps via kernel density estimation. Our experiments are based on one year's worth of mobile phone activity collected by Nokia's Mobile Data Challenge (MDC). We show that usage information can be a strong predictor of semantic place categories, allowing us to automatically annotate maps based on the behavior of the local user base.
\end{abstract}

\section{Introduction}\label{sec:intro}
Urban environments, as soon as they surpass a certain size, develop a complex structure of neighborhoods and districts that often does not correspond with the exact boundaries  chartered in official ground books and maps. As a consequence, reliably associating a city's functional neighborhoods (as opposed to the official ones) with their key functions can be a difficult endeavor requiring considerable familiarity with the place. The recent proliferation of smart phones and location-based services have given researchers and industry players access to an unprecedented amount of location data and movement traces of individual users. This information has been used in various ways to model the spatial distribution of both individuals and whole populations. Examples include modelling the spread of infectious diseases using geo-tagged tweets~\cite{sadilek2012modeling} or location-dependent prediction of crime~\cite{gerber2014predicting}. Similarly, we argue that location-based data can be used to model the structure and composition of a city. Let us begin with the specific case where we define a city in terms of its neighborhoods. They can be defined according to the distribution of resident establishments. For example, the neighborhoods around Greenwich Village and Chelsea in New York City can be classified as primarily containing entertainment and food-related venues while Brooklyn comprises primarily residential areas. One can also define neighborhoods in terms of demographic attributes such as age, gender or employment status. These can lead to the discovery of interesting, potentially interwoven, patterns which can, in turn, be used to  enhance traditional machine learning algorithms to increase the accuracy of inference tasks.

As part of the 2012 Nokia Mobile Data Challenge (MDC)~\cite{laurila2012mobile}, mobile phone usage data of approximately 200 consenting users collected over the course of a full year was made available to researchers. This data contains, among others, call logs, GPS traces, accelerometer readings, self-annotated location visits and demographic information. One of the tasks in the challenge was to accurately predict the semantic label of a place. In this paper, we revisit this task and present an automatic method for semantic place description based on kernel density estimates of geo-tagged place visits that gives fine-grained insights into the specific composition of urban landscapes without the need for manual data annotation or interaction.

The remainder of this paper is structured as follows: In Section~\ref{sec:related}, we present a brief overview of related work towards interpreting and exploiting spatial data. Section~\ref{sec:method} introduces our core methodology. After a review of the necessary geo-statistical methods and spatial clustering strategies, we discuss three main approaches used for estimating the class-conditional probability density at a given location based on its distance from other neighboring places of the same category. Section~\ref{sec:data} gives details about the Lausanne Data Collection Campaign and Nokia's Mobile Data Challenge. We describe the Semantic Place Prediction task and explore salient patterns inherent to the data. Section~\ref{sec:experiments} describes our experimental setup along with a discussion of the key findings. Section~\ref{sec:conclusion}, finally, concludes the paper and gives an outlook on future directions.

\section{Related Work}\label{sec:related}
There is a wide range of work applying geo-tagged resources for inference and retrieval tasks. 

Abrol and Khan~\cite{abrol2010twinner} present a query refinement scheme based on location information associated with relevant documents. Similarly, location information has been shown to be a useful source of evidence for Web search personalization~\cite{bennett2011inferring,leung2010personalized}, resulting in significantly improved result rankings, especially for queries with local intent. Li et al.~\cite{li2014geo} as well as Cheng et al.~\cite{cheng2014barbecue} use the location information associated with tweets in order to model the users' topical knowledge. Their methods are based on the concept of a geographical frame of reference within which they are looking for experts on topics ranging from food to entertainment. In addition to these general-purpose applications, there are numerous tailor-made approaches of using geo-tagged content for goals such as disease spread modelling~\cite{sadilek2012modeling}, crime prediction~\cite{gerber2014predicting}, disaster relief~\cite{vieweg2010microblogging}, or the prediction of Internet meme dispersion~\cite{kamath2013spatio}.

Let the above serve as a selection of use cases in which location information is used to support traditional tasks. However, the reverse is true as well and several lines of work are dedicated to automatically inferring the location of a piece of content. Cheng et al.~\cite{cheng2010you} devise a user location method on the basis of twitter activity that is able to correctly place the majority of users within a 100-mile radius of their true location. Li et al.~\cite{li2011tweet} generalize this task from determining unique locations to assigning the correct place-of-origin category to a tweet. Serdyukov et al.~\cite{serdyukov2009placing} as well as Hauff and Houben~\cite{hauff2012placing} present image location schemes, associating photos with their correct location of origin. Going beyond the mere place of origin of a piece of content, some researchers have been trying to predict also future steps the user is likely to take. Examples of this so-called next place prediction task include~\cite{gambs2012next,li2012want,noulas2012mining}.

This work is different in scope in that it tries to deliver general-purpose location descriptors by modelling the distribution of demographics and localities affiliated with a location in a spatially smooth manner. Wakamiya et al.~\cite{wakamiya2011crowd} follow a similar aim, in the sense that they, as well, rely on usage information to describe urban spaces. There are, however, some key differences. While the authors categorize whole cities into 4 types, our method is able to discover intra-city patterns of neighborhoods or streets and employs a more fine-grained location taxonomy, permitting its use for the aim of semantic place description.

\section{Methodology}\label{sec:method}
Our goal of semantic place description requires three key components: Firstly, the concept of \textit{distance} plays a key role in this work and we will need a reliable way of measuring the proximity between locations $l_i$ and $l_j$. Secondly, raw place visits $v$ observed in our data have to be spatially clustered to form meaningful regions $r$ (our neighborhoods). And finally, for each location $l$ in such a region, we need a way to smoothly distribute probability mass of all nearby observations $V_r$.

\subsection{Measuring Geo-spatial Distance}
In the course of this section we will work with a set $L$ of geographical locations $l$ specified in terms of their latitude and longitude. One of the most straight forward choices for measuring point-to-point distances is the Euclidean distance, which is simply the straight line distance between two points. While this works well for Euclidean spaces, the surface of the earth is spherical. In this case, a Great-circle distance gives a better estimate of the distance between two points. Great-circle distance describes the shortest distance $d$ between two points $l_i$ and $l_j$ on a sphere measured along the surface of the sphere. Given any two points on a sphere which are not directly opposite each other, there is a
unique great circle which passes through both of them. The two points split the great circle into two arcs. The Great-circle distance between these two points is then defined as the length of the shorter arc. The Haversine formula~\cite{robusto1957cosine}, given below, computes the Great-circle distance $d$ between two points on a sphere, given their latitude $\phi$ and longitude $\lambda$. 

\begin{center}
$d(l_i,l_j) = 2 r \; \textit{arcsin} \; \sqrt{\textit{sin}^2 (\frac{\phi_j - \phi_i}{2}) + \textit{cos}(\phi_i) \; \textit{cos}(\phi_j) \; \textit{sin}^2  (\frac{\lambda_j - \lambda_i}{2})}$
\end{center}

Where $r$ refers to the radius of the sphere (6372.8 kilometers in case of the Earth). Note that this formulation assumes that the latitude and longitude are provided in radians. It should be mentioned that the choice of this distance function over the more na\"{i}ve Euclidean distance may not make a huge difference for urban-scale problems. For the sake of generality, however, we adjust for spherical effects to allow our method to seamlessly generalize to problems at national or continental scale.

\subsection{Spatial Clustering}
With our distance metric in place, we move on to clustering spatially neighboring observations into regions. While there are readily applicable general-purpose clustering methods such as the popular K-Means algorithm, we turn towards a more geographically-informed alternative. The key difference between clustering in logical spaces (e.g., term frequency spaces) and geographical ones lies in their respective continuities. Some geographical structures such as mountains or lakes form natural boundaries on top of which many types of observations may be less likely or altogether impossible. 

As an alternative, the DBSCAN algorithm~\cite{ester1996density} views clusters as areas of high distributional density, separated by areas of low density.  Instead of modelling distances to cluster centroids, the algorithm adds points to an existing cluster if they are located within a certain radius of any points already in that cluster AND surrounded by a given threshold number of neighbors. As a result, and unlike K-means, it does not put a restriction on the shape of the clusters, making it a strong candidate for clustering places in geographic spaces. As an added benefit, it does not require any knowledge about the expected number of clusters, relying only on the permissible distance $\epsilon$ from existing clusters as well as the required number of neighbors $\eta$. Algorithm \ref{alg:dbscan} details, in pseudo code, the DBSCAN procedure.

\begin{algorithm}
    \caption{DBSCAN}
    \label{alg:dbscan}
\begin{algorithmic}
 \Function{DBScan}{$V$, $\epsilon$, $\eta$}
 \State $C = 0$
 \For {$v$ in $V$}
   \If{$v$ is visited}
     \State move to next $v$
   \EndIf
   \State mark $v$ as visited
   \State Neighbors = \Call {regionQuery}{$v$, $\epsilon$}
   \If {\Call {sizeof}{Neighbors} $<$ $\eta$}
     \State mark $v$ as NOISE
   \Else
     \State $C$ = next cluster
     \State \Call {expandCluster}{$v$, Neighbors, $C$, $\epsilon$, $\eta$}
   \EndIf
 \EndFor
 \EndFunction
 \\
 \Function {expandCluster} {$v$, Neighbors, $C$, $\epsilon$, $\eta$}
   \State add $v$ to cluster $C$
   \For {each point $v'$ in Neighbors} 
      \If {$v'$ is not visited}
         \State mark $v'$ as visited
         \State Neighbors' = \Call {regionQuery}{$v'$, $\epsilon$}
         \If {\Call {sizeof}{Neighbors'} $\geq \eta$}
            \State Neighbors = Neighbors $\cup$ Neighbors'
         \EndIf
      \EndIf
      \If {$v'$ is not yet member of any cluster}
         \State add $v'$ to cluster $C$
      \EndIf
   \EndFor
\EndFunction
\\
\Function{regionQuery}{$v$, $\epsilon$}\\
  \Return all points within $v$'s $\epsilon$-neighborhood (including $v$)
\EndFunction
\end{algorithmic}
\end{algorithm}

\subsection{Spatial Probability Mass Smoothing}
Given our distance metric and a means of grouping observations into spatially coherent regions, the final missing component in our inference scheme is a method for smoothly aggregating probability mass from previous observations to describe unseen data points. To this end, we rely on Kernel density estimation (KDE), a non-parametric method to estimate the probability density function of a random variable based on empirical observations. Let $(x_1 , x_2 , \ldots, x_n)$ be an independent and identically distributed sample drawn from some distribution with an unknown density function f. We intend to find an estimate $\hat{f}$ for this function. Its kernel density estimator is then given by Equation~\ref{eq:kde}, where $K$ is the kernel, a non-negative function that integrates to 1 and has mean 0 and $h$ is a smoothing parameter, the so-called bandwidth. 

\begin{equation}
\label{eq:kde}
\hat{f}(x) = \frac{1}{nh} \sum_{i=1}^{n}{K(\frac{x-x_i}{h})}
\end{equation}

Intuitively speaking, the estimator sums up probability mass to the final probability density estimate for a new data point from each of the existing data points. There are numerous kernel functions that can be applied here. After having experimented with a number of popular choices (uniform, triangular, biweight, triweight, Epanechnikov, exponential and Gaussian), we obtained best results for the Gaussian kernel given in Equation~\ref{eq:kernel}.

\begin{equation}
\label{eq:kernel}
K(x) = \frac{1}{\sqrt{2 \pi}} e^{-\frac{1}{2}x^2}
\end{equation}

Following the general KDE formulation, the contribution of each data point depends on the bandwidth parameter $h$, but is generally higher for points close to the sampled data point. The concrete choice of the bandwidth has a strong influence on the final density estimate. If we choose small values of $h$, the contributions of the points closest to the probe point are magnified. In this case the estimator is said to be ``\textit{undersmoothed}''. If we choose too large a bandwidth on the other hand, then we do not adequately account for the local context. The resulting estimator is ``\textit{oversmoothed}''. In practice, cross-validation methods for bandwidth selection have been shown to perform well across a wide range of data sets~\cite{hall1992smoothed}. As an alternative to such uniform bandwidths that often fail to adapt appropriately in regions of relatively high or low densities, there are methods for dynamically setting $h$ based on the local probe context. To that end, we rely on balloon bandwidth estimators~\cite{terrell1992variable} as given in Equation~\ref{eq:balloon}. Under this framework, a simple way to find the bandwidth $h(x)$ for a probe point is to set it to the distance between $x$ and the $k$-th sample point. In practice, best results were achieved for settings of $k = 15$.

\begin{equation}
\label{eq:balloon}
\hat{f}(x) = \frac{1}{nh(x)} \sum_{i=1}^{n}{K(\frac{x-x_i}{h(x)})}
\end{equation}

\subsection{Spatial Kernel Discriminant Analysis}
With all our individual components in place, the overall processing becomes a simple pipeline: We first group our observations according to their class labels. We train an estimator for each of the labels. Given a new place with its latitude and longitude, we can now determine if it lies in a residential area of the city or in the shopping center based on the respective classes' probability density estimates. Each of the $k$ classes is associated with a density function $\hat{f}_j$ and a prior probability $p_j$ . We wish to associate a new location x with exactly one of the $k$ categories. Following the Kernel Discriminant Rule, we arrive at the maximum likelihood formulation in Equation~\ref{eq:kda}.

\begin{equation}
\label{eq:kda}
\textit{category} \; (x) = \textit{arg max} \hat{f}_j (x)
\end{equation}

The use of geo-spatial clustering (e.g., DBSCAN), at this point, is optional. We can either distribute probability mass globally across all observations or first group our observations into geographically cohesive regions and subsequently run KDE only within the region into which the new sample location falls. Section~\ref{sec:experiments} will give a detailed look into the individual classification performance related to these and several other design decisions. 

\section{Dataset}\label{sec:data}
This paper relies on the data provided by the Nokia Mobile Data Challenge (MDC) held in 2012~\cite{laurila2012mobile}. Between the years of 2009 and 2011, approximately 200 volunteers from the Lake Geneva region in Switzerland were equipped with okia N95 smart phones and their mobility and usage patterns were recorded over the course of more than one year. During this period, a wide range of signals were logged. Table~\ref{tab:data} gives an overview of the various types of information provided in the dataset along with their available frequency of measurement.

\begin{table}
\label{tab:data}
\caption{Data Collected from LDCC}
\centering{}
\begin{tabular}{|c|r|}
\hline
Data type & Quantity  \\
\hline
\hline
Calls & 240,227  \\ 
\hline 
Text messages & 175,832 \\ 
\hline 
Phone book entries & 45,928  \\
\hline 
Location points & 26,152,673  \\
\hline 
Unique cell towers & 99,166 \\
\hline 
Accelerometer samples & 1,273,333\\
\hline 
WiFi observations & 31,013,270  \\
\hline 
Unique WiFi access points & 560,441\\
\hline 
Unique Bluetooth devices & 498,593\\
\hline 
Bluetooth observations & 38,259,550\\
\hline 
\end{tabular}  
\end{table} 

\subsection{Demographics}
The campaign's population of 191 participants is 38\% male and 62\% female. Due to active recruiting at colleges and universities, the participant base mainly consists of younger people, with the majority of the population aged between 22-33 years old. Alongside the actual phone usage data, the following participant demographics were collected via a survey:
\begin{itemize}
\item Age of the participant
\item Gender of the participant
\item Employment status of the participant
\item Who pays the phone bill of the participant
\item Number of members in the family of the participant
\end{itemize}

For a more detailed overview of participant demographics, please refer to the original dataset description~\cite{laurila2012mobile}.

\subsection{Semantic Place Labels}
As part of the survey, each user assigned semantic labels for select, frequently visited, places. Candidate places were inferred automatically and then presented to the users for labeling. Inferring the semantic role of places is a relevant problem with applications in many fields, including, for example, mobile computing~\cite{isaacman2011identifying}. The semantic place labels form the basis of the MDC's semantic place prediction task. Each place is represented by a history of visits along with some basic contextual information such as time of the day, accelerometer readings, etc.\ for each visit. Each place is annotated with one of 10 pre-defined semantic labels. There are a total of 493 semantically annotated place visits. Since this labelling step was an optional procedure, not all participants contributed annotations. Table~\ref{tab:places} enumerates the labels and describes their distribution in the dataset.

\begin{table}
\label{tab:places}
\caption{Frequency of Semantic Place Labels}
\centering
\begin{tabular}{|c|c|c|}
\hline
Semantic Label & Frequency (abs) & Frequency (rel)\\
\hline 
\hline
Bar;Restaurant & 14 & 2.9\%\\
\hline
Outdoor Sports & 19 & 3.9\%\\ 
\hline
Indoor Sports & 31 & 6.4\%\\ 
\hline
Home & 124 & 25.4\%\\
\hline
Home of a Friend & 76 & 15.6\%\\
\hline
Transport Related & 34 & 7.0\%\\
\hline
Work & 142 & 29.1\%\\
\hline
Shop & 24 & 4.9\%\\
\hline
Holidays Resort & 10 & 2.1\%\\
\hline
Work of a Friend & 14 & 2.9\%\\
\hline
\end{tabular}  
\end{table} 

\subsection{Inferring Incomplete Geo-Coordinates}
One of the design goals of the MDC was to adequately handle the trade off between collecting comprehensive data and judicious power consumption. To that end, the organizers ensured non-intrusive device usage during the day, without the need to charge the phone until the night~\cite{kiukkonen2010towards}. This was achieved by having the data collection software tune sampling rates based on the battery levels and activity profiles.

As a consequence, in the MDC dataset, places are not explicitly annotated with location information. Since this type of information, however, represents a core component of our method, we made use of the available affiliated information in order to accurately infer the location of each place visit. 

GPS traces and WiFi information recorded as part of the study. Whenever the user is at a new location, the software first tries to detect if there are WiFi connections available nearby. If there are and if they are already present in a local table on the phone, the software just records the timestamp when the connection was made. If a new WiFi connection is discovered, the software records the location of the hotspot and its identifier along with the timestamp in the local database. Only if no WiFi hotspots at all are found, the phone activates the GPS and logs the GPS traces along with timestamps.

All place visits in the MDC dataset are annotated with start and end timestamps. Thisenables us to accurately infer the latitude and longitude of a visit. For all visits to a place $p$, we try to record the available WiFi hotspots. We then take the arithmetic mean of the latitudes and longitudes of all the recorded positions and set the resulting centroid as the geo-coordinate of the place. The location of 336 out of 493 places was deduced in this way. For those places which had no recorded WiFi hotspots for any of the visits, we default to raw GPS traces. Just like for WiFi connections, we record all GPS samples recorded between the start and end of all visits made to place $p$ and again take their arithmetic mean. The location of 152 out of 493 places was deduced in this way. This leaves us with a total of 488 (336 + 152) visits with known location. The remaining 5 cases in which no reliable location information could be inferred, were excluded from all further experiments.

\section{Experiments}\label{sec:experiments}

As mentioned earlier, the empirical performance evaluation of our method is based on the previously described Nokia Mobile Data Challenge (MDC) corpus. All compared methods are trained and evaluated in a stratified 10-fold cross validation scheme and performance if evaluated in terms of prediction accuracy.

\begin{figure}
\centering
\includegraphics[width=0.7\textwidth]{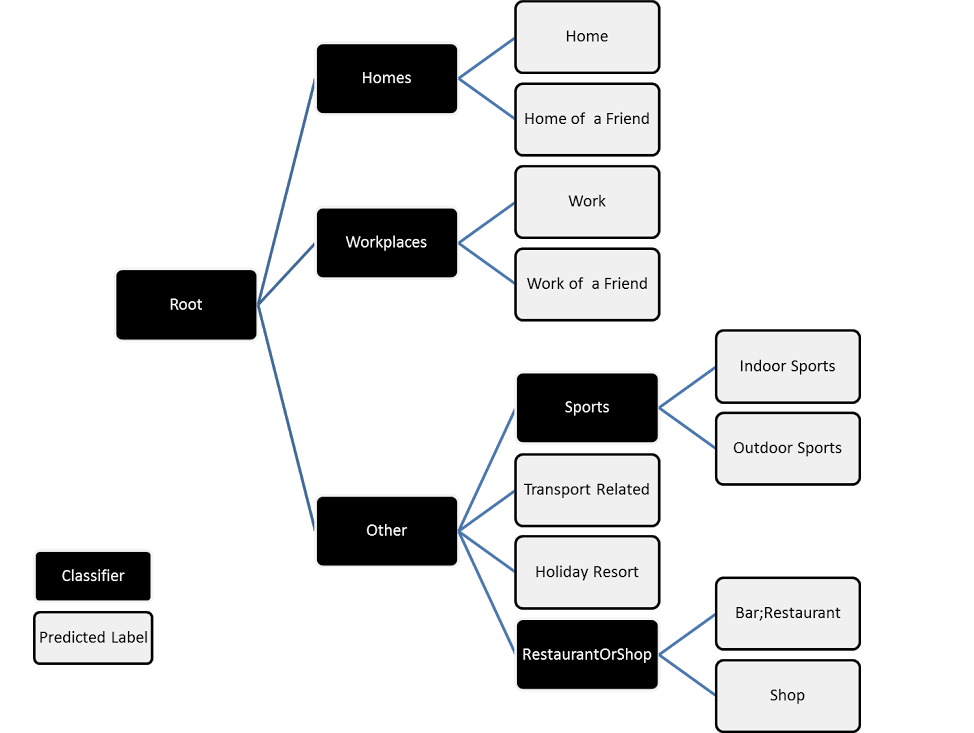}
\caption{Multi-step Classification for Semantic Place Prediction.}\label{fig:classifier}
\end{figure}

Our method is based on the work of Huang et al.~\cite{huang2012mining}, whose approach produced the overall best results among the entries of the original MDC competition and relies on a multi-level classification scheme. A hierarchical ensemble of SVM classifiers first divides visits into the high-level categories ``\textit{home}'', ``\textit{work}'' and ``\textit{other}'' and follows up with a series of dedicated low-level classifiers. The entire process is schematically described in Figure~\ref{fig:classifier}. 

Their method represents visits in terms of 54 features derived from various mobile phone sensor readings. Examples include accelerometer readings, date, time, application and phone usage statistics and many others. Please refer to their original publication for a comprehensive overview of all features.

Since none of the features considered by Huang et al.~\cite{huang2012mining} or any of their best-performing competitors directly model geo-location information, we extend their method by including prior probabilities for each class. For each of the ten semantic categories, we compute the class-conditional probability of observing a place of the respective class at the given location. The probabilities are obtained via our KDE scheme described in Section~\ref{sec:method} and take the role of priors describing the location-specific distribution of places. To better understand and interpret the results of our comparison, we further measure the performance of our method's individual components as well as several intuitive baselines. Table~\ref{tab:results} compares the respective performance of random (R) and dominant class (DC) baselines to pure KDE with fixed (KDE-F) or adaptive bandwidth (KDE-A), using DBSCAN prior to KDE-based probability smoothing (KDE-A + DBSCAN), as well as Huang et al.'s multi-level classifier ensemble (MLC) and, finally, our enhanced version thereof (MLC + KDE-A + DBSCAN).

\begin{table}
  \centering
  \caption{Semantic Place Prediction Results}
  \label{tab:results}
  \begin{tabular}{|l|c|}
    \hline
    Method & Accuracy(\%)\\
    \hline
    \hline
    Random prediction (R) & 10\\
    \hline
    Dominant class prediction (DC) & 20.6\\
    \hline
    KDE with fixed bandwidth (KDE-F) & 33.92\\
    \hline
    KDE with adaptive bandwidth (KDE-A) & 35.55\\
    \hline
    KDE-A + DBSCAN & 38.04\\
    \hline
    Multi-level Classification (MLC) & 63.73\\
    \hline
    MLC + KDE-A + DBSCAN  & 64.55*\\
    \hline
  \end{tabular}
\end{table}

We can see that our geo-spatial semantic descriptors, even in isolation, taking as input no information beyond the visit's location, deliver a solid classification performance that lies significantly above random or educated guessing. We further note that both adaptive bandwidth estimation as well as geo-spatial clustering prior to classification appear to hold considerable merit. When we finally combine our geo-spatial descriptors with the wealth of other available sources of information, they introduce a significant performance increase with respect to the strongest competing entry to the original Mobile data Challenge. Statistical significance of our findings was measured by means of a Wilcoxon Signed-Rank test at $\alpha < 0.05$-level.

\section{Conclusion}\label{sec:conclusion}
In this paper, we described a general-purpose framework for semantic place description on the basis of smoothed spatial probability density estimates. We gave an overview of supporting techniques such as adaptive bandwidth estimation and geo-spatial clustering prior to the actual classification, finding that both show significant positive impact on the descriptor's performance. In a final empirical investigation based on the well-known Nokia Mobile Data Challenge (MDC), we demonstrated how the winning entries to the original contest can be significantly improved by the inclusion of our method.

There are multiple promising research directions that we are excited to pursue in the future. (1) In this work, we exclusively focused on describing the spatial distribution of semantic class labels. There are, however, many other sources of evidence, such as demographic attributes that lend themselves to similar treatment. In a preliminary set of experiments we observed further solid gains of approximately 1\% absolute performance obtained from the use of spatial demographics. (2) Currently, we have been exclusively drawing from MDC-internal data, to show the fairest and most direct performance comparison with previous work. In the future, the broad availability of location-based services such as Google Places, Yelp or Foursquare should be exploited to obtain a better and more dense coverage of training labels. (3) Finally, while the current paper employs our geo-spatial descriptors solely for the goal of classification, there are many exciting applications such as city-map enhancement or even generation that we will explore in the future.

\bibliographystyle{plain}
\bibliography{ref}  

\end{document}